\documentclass[twocolappendix,appendixfloats]{emulateapj}

\usepackage[colorlinks = true, linkcolor = blue, urlcolor  = blue, citecolor = blue, anchorcolor = blue]{hyperref}
\usepackage{hyperref}
\usepackage{apjfonts}
\usepackage{rotating}
\usepackage{amsmath}
\usepackage{color}
\usepackage{color}
\usepackage{CJK}
\usepackage{array}

\newcommand{\pivec}{\mbox{\boldmath $\pi$}}
\newcommand{\muvec}{\mbox{\boldmath $\mu$}}

\newcommand{\thetae}{\theta_{\rm E}}
\newcommand{\pie}{\pi_{\rm E}}
\newcommand{\pien}{\pi_{{\rm E},N}}
\newcommand{\piee}{\pi_{{\rm E},E}}

\definecolor{darkbrown}{RGB}{139,69,19}

\shorttitle{KMT-2016-BLG-2052}
\shortauthors{Han et al.}

\begin{document}

\title{KMT-2016-BLG-2052L: Microlensing Binary Composed of 
M Dwarfs Revealed from a Very Long Time-scale Event}

\author{
Cheongho Han\altaffilmark{01},
Youn Kil Jung\altaffilmark{02,201}, 
Yossi Shvartzvald\altaffilmark{13,201,202},\\ 
and\\ 
Michael D.~Albrow\altaffilmark{04}, 
Sun-Ju Chung\altaffilmark{02,05}, 
Andrew Gould\altaffilmark{02,06,07}, 
Kyu-Ha Hwang\altaffilmark{02}, 
Doeon Kim\altaffilmark{01}, 
Chung-Uk Lee\altaffilmark{02}, 
Woong-Tae Kim\altaffilmark{08}, 
Hyoun-Woo Kim\altaffilmark{02}, 
Yoon-Hyun Ryu\altaffilmark{02}, 
In-Gu Shin\altaffilmark{03}, 
Jennifer C.~Yee\altaffilmark{03}, 
Chun-Hwey Kim\altaffilmark{09}, 
Sang-Mok Cha\altaffilmark{02,10}, 
Seung-Lee Kim\altaffilmark{02,05}, 
Dong-Jin Kim\altaffilmark{02}, 
Dong-Joo Lee\altaffilmark{02}, 
Yongseok Lee\altaffilmark{02,10}, 
Byeong-Gon Park\altaffilmark{02,05}, 
Richard W.~Pogge\altaffilmark{06}   \\
(The KMTNet Collaboration), \\
Charles Beichman\altaffilmark{12}, 
Geoff Bryden\altaffilmark{11}, 
Sebastiano Calchi Novati\altaffilmark{13}, 
B.~Scott Gaudi\altaffilmark{06}, 
Calen B.~Henderson\altaffilmark{12}, 
Matthew T.~P.\altaffilmark{06},
Savannah R.~Jacklin\altaffilmark{15} \\
(The UKIRT Microlensing Team)\\
}

\email{cheongho@astroph.chungbuk.ac.kr}

\altaffiltext{01} {Department of Physics, Chungbuk National University, Cheongju 28644, Republic of Korea} 
\altaffiltext{02} {Korea Astronomy and Space Science Institute, Daejon 34055, Republic of Korea} 
\altaffiltext{03} {Harvard-Smithsonian Center for Astrophysics, 60 Garden St., MS-15 Cambridge, MA, 02138, USA} 
\altaffiltext{04} {University of Canterbury, Department of Physics and Astronomy, Private Bag 4800, Christchurch 8020, New Zealand} 
\altaffiltext{05} {Korea University of Science and Technology, 217 Gajeong-ro, Yuseong-gu, Daejeon 34113, Republic of Korea} 
\altaffiltext{06} {Department of Astronomy, Ohio State University, 140 W.\ 18th Ave., Columbus, OH 43210, USA} 
\altaffiltext{07} {Max Planck Institute for Astronomy, K\"onigstuhl 17, D-69117 Heidelberg, Germany} 
\altaffiltext{08} {Department of Physics \& Astronomy, Seoul National University, Seoul 151-742, Republic of Korea} 
\altaffiltext{09} {Department of Astronomy and Space Science, Chungbuk National University, Cheongju 28644, Republic of Korea} 
\altaffiltext{10} {School of Space Research, Kyung Hee University, Yongin 17104, Republic of Korea} 
\altaffiltext{11} {Jet Propulsion Laboratory, California Institute of Technology, 4800 Oak Grove Drive, Pasadena, CA 91109, USA} 
\altaffiltext{12} {IPAC/NExScI, Mail Code 100-22, Caltech, 1200 E. California Blvd., Pasadena, CA 91125} 
\altaffiltext{13} {IPAC, Mail Code 100-22, Caltech, 1200 E. California Blvd., Pasadena, CA 91125} 
\altaffiltext{14} {Kepler \& K2 Missions, NASA Ames Research Center, PO Box 1,M/S 244-30, Moffett Field, CA 94035} 
\altaffiltext{15} {Vanderbilt University, Department of Physics \& Astronomy, Nashville, TN 37235, USA} 
\altaffiltext{201}{The KMTNet Collaboration} 
\altaffiltext{202}{The UKIRT Microlensing Team}

\begin{abstract}
We present the analysis of a binary microlensing event KMT-2016-BLG-2052, for which 
the lensing-induced brightening of the source star lasted for 2 seasons.  We determine 
the lens mass from the combined measurements of the microlens parallax $\pie$ and angular 
Einstein radius $\thetae$.  The measured mass indicates that the lens is a binary composed 
of M dwarfs with  masses of $M_1\sim 0.34~M_\odot$ and $M_2\sim 0.17~M_\odot$.  The measured 
relative lens-source proper motion of $\mu\sim 3.9~{\rm mas}~{\rm yr}^{-1}$ is smaller than 
$\sim 5~{\rm mas}~{\rm yr}^{-1}$ of typical Galactic lensing events, while the estimated 
angular Einstein radius of $\thetae\sim 1.2~{\rm mas}$ is substantially greater than the 
typical value of $\sim 0.5~{\rm mas}$.  Therefore, it turns out that the long time scale 
of the event is caused by the combination of the slow $\mu$ and large $\thetae$ rather 
than the heavy mass of the lens.  From the simulation of Galactic lensing events with 
very long time scales ($t_{\rm E}\gtrsim 100$ days), we find that the probabilities that 
long time-scale events are produced by lenses with masses $\geq 1.0~M_\odot$ and 
$\geq 3.0~M_\odot$ are $\sim 19\%$ and 2.6\%, respectively, indicating that events produced 
by heavy lenses comprise a minor fraction of long time-scale events.  The results indicate 
that it is essential to determine lens masses by measuring both $\pie$ and $\thetae$ in 
order to firmly identify heavy stellar remnants such as neutron stars and black holes.
\end{abstract}


\section{Introduction}\label{section:one}

From dozens per year when the first-generation microlensing experiments \citep{Alcock1993, 
Udalski1993, Aubourg1993} were conducted, the detection rate of microlensing events has 
greatly increased and currently more than 2500 microlensing events are annually detected
\citep{Udalski2015, Bond2001, Kim2018b}.  
The dramatic increase of the event rate became possible by various factors including the 
development of advanced event finding algorithms, the increased observational cadence thanks 
to upgraded instruments, and the addition of new surveys.  As the event rate increases, 
the scientific scope of microlensing has also expanded from the original use of detecting 
Galactic dark matter in the form of massive compact halo objects \citep{Paczynski1986} 
into various fields including extrasolar planet searches \citep{Mao1991,Gould1992b}.

A small fraction of microlensing events last for very long durations. Such long time-scale 
events are of scientific importance for various reasons. First, lenses of these events are 
candidates of heavy stellar remnants such as neutron stars (NSs) and black holes (BHs)
\citep{Shvartzvald2015, Wyrzykowski2016}.  The 
event time scale, which is defined as the time for the source to cross the angular Einstein 
radius $\thetae$ of the lens, is related to the physical parameters of the lens system by
\begin{equation}
t_{\rm E}={\thetae\over \mu} = 
{\sqrt{\kappa M \pi_{\rm rel}} \over \mu}, 
\label{eq1}
\end{equation}
where $M$ is the lens mass, $\mu$ is the relative lens-source proper motion, 
$\kappa=4G/(c^2{\rm au})$, $\pi_{\rm rel}=\pi_{\rm L}-\pi_{\rm S}={\rm au}
(D_{\rm L}^{-1}-D_{\rm S}^{-1})$ represents the relative lens-source parallax, and $D_{\rm L}$ 
and $D_{\rm S}$ denote the distances to the lens and source, respectively. Because the time scale 
is proportional to the square root of the lens mass, very long time-scale events are more likely 
to be produced by heavy lenses.  Second, the chance to measure a microlens parallax $\pie$ is 
high for long time-scale events.  As an event time scale approaches or exceeds the orbital 
period of Earth, i.e., 1~yr, the relative lens-source motion departs from rectilinear due 
to Earth's orbital motion.  This induces long-term deviations in lensing light curves, 
microlens-parallax effects, and the analysis of the deviation yields $\pie$ \citep{Gould1992a}. 
The microlens parallax is related to the lens mass and distance by
\begin{equation}
M={\thetae\over \kappa \pie}
\label{eq2}
\end{equation}
and
\begin{equation}
D_{\rm L}={{\rm au} \over \pie\thetae+\pi_{\rm S}},
\label{eq3}
\end{equation}
respectively \citep{Gould2000b}.  Therefore, the physical lens parameters can be significantly 
better defined with the additional constraint of the microlens parallax \citep{Han1995}. Third, 
long time-scale events produced by binary lenses are especially important because one can 
additionally measure the angular Einstein radius. This is because binary-lens events usually 
produce caustic-crossing features in lensing light curves. This part of the light curve is 
affected by finite-source effects, and the analysis of the deviation enables one to measure 
the Einstein radius. With the measurement of both $\pie$ and $\thetae$, the lens mass can be 
uniquely determined and the nature of the lens can be revealed.

In this work, we present the analysis of a binary microlensing event KMT-2016-BLG-2052. For 
the event, the lensing-induced magnification of the source flux lasted for two years from the 
beginning of 2016 bulge season until the end of 2017 season. The light curve of the event also 
exhibits a caustic-crossing feature that was densely resolved.  We characterize the lens by 
estimating the mass from the simultaneous measurements of $\pie$ and $\thetae$.

\begin{figure*}
\epsscale{0.95}
\plotone{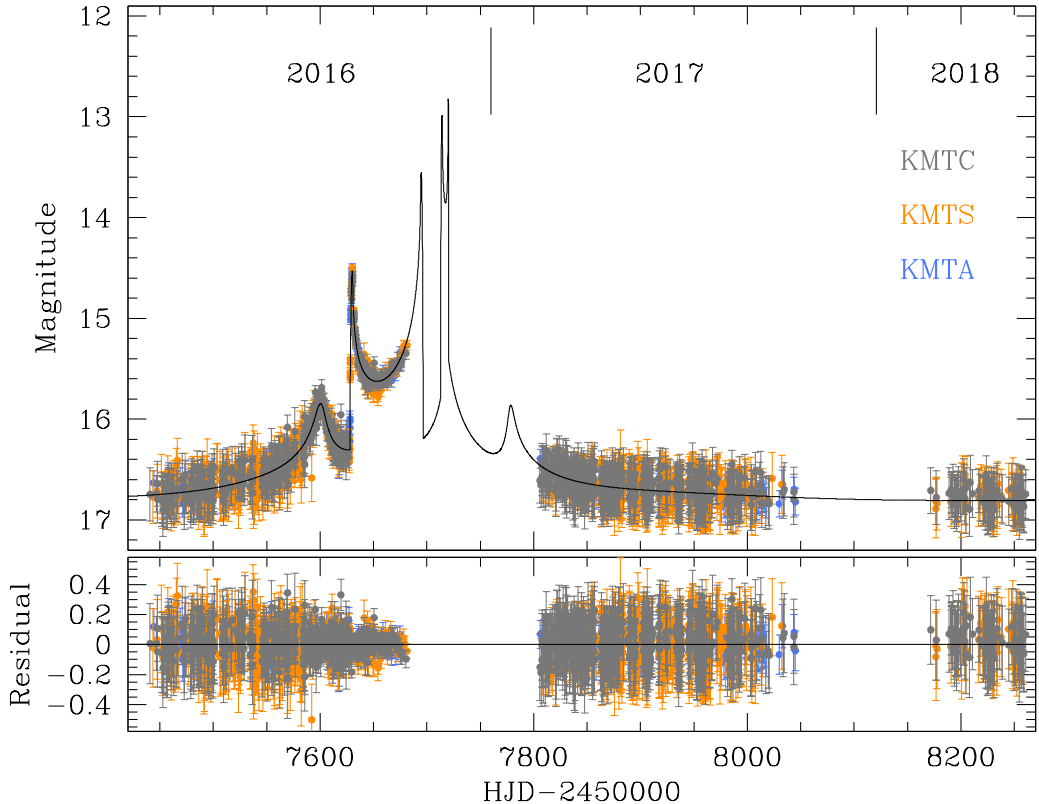}
\caption{
Light curve of KMT-2016-BLG-2052. The curve superposed
on the data points represents the binary-lensing model.
The lower panel shows the residual from the model.\\
\smallskip
}
\label{fig:one}
\end{figure*}

\section{Observation and Data}\label{section2}

The lensing event KMT-2016-BLG-2052 occurred on a star located toward the Galactic bulge field with 
equatorial coordinates $({\rm RA},{\rm dec})_{\rm J2000}=$(17:41:19.50, -27:40:19.67), which 
corresponds to Galactic coordinates $(l,b)=(0.58^\circ,1.47^\circ)$. Due to the closeness to 
the Galactic center, the source star was heavily extincted by dust.

The event was identified by applying the Event Finder algorithm \citep{Kim2018a,Kim2018b} 
to the 2016 season data acquired by Korea Microlensing Telescope Network (KMTNet) survey 
\citep{Kim2016}. The survey uses three identical 1.6~m telescopes that are globally located 
at the Cerro Tololo Interamerican Observatory in Chile, the South African Astronomical 
Observatory in South Africa, and the Siding Spring Observatory in Australia. We designates 
the individual KMTNet telescopes as KMTC, KMTS, and KMTA, respectively. 
The observations were conducted mostly in $I$ band with occasional $V$-band observations for 
the source color measurement. The source is located in the BLG15 field for which observations 
were conducted at one-hour cadence.  The data were reduced using the pySIS photometry software 
package \citep{Albrow2009} that is developed on the basis of the Difference Image Analysis 
technique \citep{Alard1998, Wozniak2000}.  For the KMTC data set, additional photometry is 
conducted using the software package DoPHOT \citep{Schechter1993} for the construction of 
color-magnitude diagram and the measurement of the source color.  The data sets used in the 
analysis are composed of 1168, 1132, and 410 points collected from the KMTC, KMTS, and KMTA 
observations, respectively.

\begin{figure}
\includegraphics[width=\columnwidth]{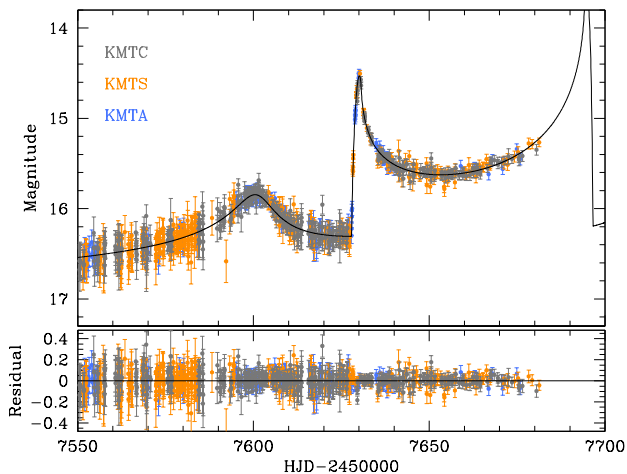}
\caption{
Enlarged view of the light curve covering the observed portion of the features induced by 
the binary caustic.\\
\smallskip
}
\label{fig:two}
\end{figure}

The event was also observed in 2015 and 2017 seasons using the 3.8~m United Kingdom Infrared 
Telescope \citep[UKIRT Micro-lensing survey:][]{Shvartzvald2017}.  UKIRT observations were 
conducted in $H$ band and aperture photometry is used for reduction. 
The data were used for the source color measurement.  The UKIRT data 
set is composed of 142 and 75 points taken in 2015 and 2017 seasons, respectively.

In Figure~\ref{fig:one}, we present  the light curve of the event. The most important 
characteristics of the event is its long duration.  The lensing-induced magnification of the 
source flux started from the beginning of 2016 bulge season and continued until the end of 
the season.  Due to the scientific importance of a long time-scale caustic-crossing binary 
lens event, we have incorporated additional data from the 2017 season.  Surprisingly, the 
event continued until the end of 2017 season.  The light curve is featured by a bump centered 
at ${\rm HJD}^\prime={\rm HJD}-2450000\sim 7600$ and a sharp spike at ${\rm HJD}^\prime\sim 7630$.  
See Figure~\ref{fig:two}, where we present the enlarged view around the features.  These bump 
and spike features are produced when a source approaches close to the cusp and passes over the 
fold of a caustic formed by a binary lens, respectively.  Binary-lens caustics form closed 
curves and thus caustic crossings occur in pairs and the light curve between the caustic 
crossings is characterized by a ``U''-shape trough.  From the partial U-shape feature observed 
during $7630 \lesssim {\rm HJD}^\prime \lesssim 7680$, it is very likely that the second caustic 
crossing (and, possibly, additional caustic-related features) occurred during the 4-month period 
when the bulge field was not observed as it passed behind the Sun.  Because the event did not 
return to the baseline until the end of 2017 season, we incorporate additional data collected 
during the 2018 season in the analysis for the secure baseline measurement.

\section{Analysis}\label{section3}

Because the bump and spike are characteristic features of binary lens events, we conduct modeling 
of the observed light curve based on the binary-lens interpretation. Under the assumption that 
there is no acceleration in the relative lens-source motion, a binary lensing light curve is 
described by 7 principal parameters. Four of these parameters describe the lens-source approach 
including the time of the closest source approach to a reference position of the lens, $t_0$, 
the source-reference separation at that time, $u_0$, the event time scale $t_{\rm E}$, and the 
angle between the source trajectory and the binary axis, $\alpha$ (source trajectory angle).  
For the reference lens position, we use the center of mass.  Two parameters $s$ and $q$ represent 
the projected separation (normalized to $\thetae$) and mass ratio between the binary lens 
components, respectively. The last parameters $\rho$, which is defined as the ratio of the 
angular source radius $\theta_*$ to $\thetae$ (normalized source radius), is needed to account 
for finite-source effects which cause deviations in lensing light curves when the source crosses 
over or approaches close to caustics.

We begin modeling the light curve with the principal binary-lensing parameters under the assumption 
that the relative lens-source motion is rectilinear. The modeling is conducted in two steps. In the 
first step, we conduct a grid search for $s$ and $q$, while the other parameters are searched for 
using a downhill approach based on the Markov Chain Monte Carlo (MCMC) method. This preliminary 
search yields a $\chi^2$ map in the $(\log s, \log q)$ plane from which we identify local minima 
and possible degenerate solutions. Because the nature of the lens is not known in advance, we inspect 
a wide range of binary separations and mass ratios. The inspected ranges are $-1.0 < \log s <1.0$ 
and $-5.0 < \log q <1.0$.  For the local minima found from this preliminary search, we then refine 
the solutions by allowing all lensing parameters to vary. From this preliminary modeling, we identify 
a solution with $s\sim 1.4$ and $q\sim 0.26$.  The model describes the overall light curve.  However, it 
leaves substantial residuals, especially around the main features of the bump and caustic-crossing spike.

Because the event lasted for $\sim 2$ years, the assumption of a rectilinear relative lens-source 
motion may not be valid due to either the orbital motion of the observer, i.e., Earth, or of the binary 
lens. We, therefore, test whether the fit improves with the consideration of the higher-order effects 
caused by the orbital motions of Earth (microlens-parallax effect) and the lens (lens-orbital 
effect). Modeling the light curve considering the microlens-parallax effect requires two additional 
parameters $\pien$ and $\piee$, which denote the north and east components of the microlens-parallax 
vector $\pivec_{\rm E}=(\pi_{\rm rel}/\thetae)(\muvec/\mu)$, respectively.  Consideration of the 
lens-orbital motion also requires to include additional parameters. Under the approximation that 
the positional changes of the lens components are small during the event, the effect is described 
by the two parameters of $ds/dt$ and $d\alpha/dt$, which denote the change rates of the binary separation 
and the orientation angle of the binary axis relative to the source trajectory,
respectively. When microlens-parallax effects are considered, it is 
known that there exist a pair of degenerate solutions with $u_0>0$ and $u_0<0$ due to mirror symmetry 
of the source trajectory with respect to the binary axis: ``ecliptic degeneracy'' \citep{Smith2003, 
Skowron2011}. Therefore, we test the degeneracy whenever microlens-parallax effects 
are considered.

\begin{figure}
\includegraphics[width=\columnwidth]{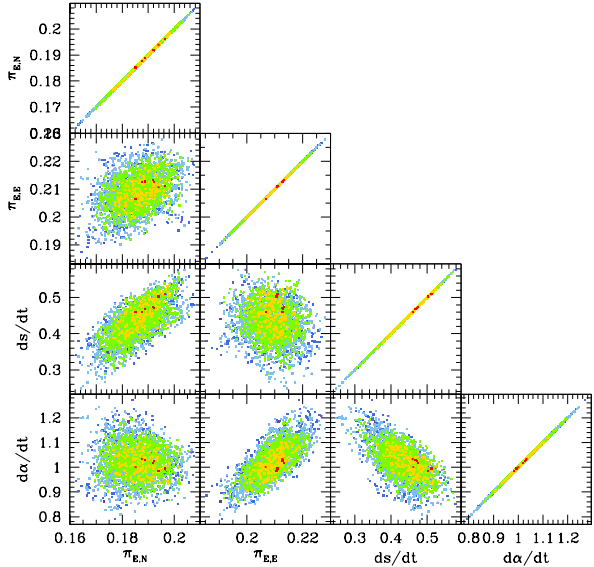}
\caption{
Distributions of $\Delta\chi^2$ in the planes of the higher-order lensing parameter pairs. The color 
coding indicates points in the MCMC chain within 1$\sigma$ (red), 2$\sigma$ (yellow), 3$\sigma$ 
(green), 4$\sigma$ (cyan), and 5$\sigma$ (blue) with respect to the best-fit value.
The distributions are for the $u_0>0$ solution.\\
\smallskip
}
\label{fig:three}
\end{figure}

\begin{table}
\centering
\caption{Comparison of Models.}
\begin{tabular}{lccc}
\hline\hline
\multicolumn{2}{c}{\ \ \ \ \ \ Model\ \ \ \ \ \ } &
\multicolumn{1}{c}{\ \ \ \ \ \ \ \ \ \ \ \  $\chi^2$\ \ \ \ \ \ \ \ \ \ \ \ } \\
\hline
Standard           &              &  3159.7  \\
Orbit              &              &  2721.8  \\
Parallax           &  ($u_0>0$)   &  2712.6  \\
                   &  ($u_0<0$)   &  2712.7  \\
Orbit+Parallax     &  ($u_0>0$)   &  2675.2  \\
                   &  ($u_0<0$)   &  2691.2  \\                   
\hline
\end{tabular}
\label{table:one}
\end{table}

\begin{figure}
\includegraphics[width=\columnwidth]{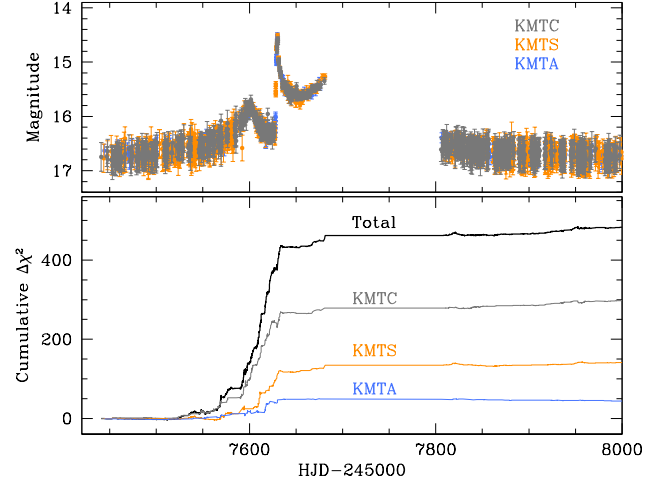}
\caption{
Cumulative distribution of $\Delta\chi^2$ between the models with and without the higher-order 
effects. The thick black solid curve is for the whole data and the thin curves with different 
colors are for the individual data sets. The light curve in the upper panel is presented to 
show the region of fit improvement.\\
\smallskip
}
\label{fig:four}
\end{figure}

In Table~\ref{table:one}, we list the $\chi^2$ values of the tested models. The ``standard'' model 
designates the solution obtained under the assumption of a rectilinear relative lens-source motion. 
In the ``orbit'' and ``parallax'' models, we separately consider the lens-orbital and microlens-parallax 
effects, respectively. In the ``orbit+parallax'' model, we simultaneously consider both higher-order 
effects. We find that higher-order effects greatly improve the fits.  As measured by the difference 
in $\chi^2$ values, the improvement is $\Delta\chi^2\sim 434$ and $\sim 447$ with respect to the 
standard model when the lens-orbital and microlens-parallax effects are separately considered, 
respectively. When both higher-order effects are simultaneously considered, the fit improves by 
$\Delta\chi^2\sim 485$.  It is found that the degeneracy between $u_0>0$ and $u_0<0$ solutions is 
moderately severe with $\Delta\chi^2\sim 16$.  Considering that the improvement by the individual 
higher-order effects are similar, it is likely that both the microlens-parallax and lens-orbital 
effects are important to precisely describe the observed light curve.  It is known that both higher-order 
effects cause qualitatively similar deviations in lensing light curves \citep{Batista2011, Skowron2011, Han2016}.  
In Figure~\ref{fig:three}, we present the $\Delta\chi^2$ distributions of MCMC points in the planes of the 
higher-order lensing parameter pairs to show the correlations between the higher-order lensing parameters.  
It shows that the $\pien$--$ds/dt$ and $\piee$--$d\alpha/dt$ parameter pairs are closely correlated.  
To check the region of the fit improvement by the higher-order effects, in Figure~\ref{fig:four}, we plot 
the cumulative distribution of $\Delta\chi^2$ between the models with and without the higher-order effects. 
For all data sets, the fit improves throughout the event.

In Table~\ref{table:two}, we present the lensing parameters of the best-fit solutions. Because 
the degeneracy between the $u_0>0$ and $u_0<0$ solutions is moderately severe, we present the 
lensing parameters of both solutions. From the lensing parameters, it is found that the event 
was produced by a binary with a mass ratio $q\sim 0.5$ and a projected separation very close 
to the Einstein radius, i.e., $s\sim 1.0$.  As anticipated, the measured event time scale, 
$t_{\rm E}\sim 112$ days, is very long.  We note that the lensing parameters of the $u_0>0$ 
and $u_0<0$ solutions are roughly in the relation 
$(u_0, \alpha, \pien, d\alpha/dt) \leftrightarrow -(u_0, \alpha, \pien, d\alpha/dt)$ due to 
the mirror symmetry of the lens system configuration \citep{Skowron2011}.

\begin{table}
\centering
\caption{Best-fit lensing parameters.}
\begin{tabular}{lcc}
\hline\hline
\multicolumn{1}{c}{Parameter} &
\multicolumn{1}{c}{$u_0>0$}   &
\multicolumn{1}{c}{$u_0<0$}   \\
\hline
$t_0$ (HJD$^\prime$)           &  7709.427 $\pm$ 0.823  &   7709.939 $\pm$ 0.592    \\
$u_0$                          &  0.174 $\pm$ 0.009     &   -0.204 $\pm$ 0.006      \\
$t_{\rm E}$ (days)             &  111.53 $\pm$ 0.78     &   112.57 $\pm$ 0.45       \\
$s$                            &  0.957 $\pm$ 0.005     &   0.958 $\pm$ 0.001       \\
$q$                            &  0.507 $\pm$ 0.011     &   0.525 $\pm$ 0.011       \\
$\alpha$ (rad)                 &  4.258  $\pm$ 0.022    &   -4.206 $\pm$ 0.011      \\
$\rho$ ($10^{-3}$)             &  2.71 $\pm$ 0.07       &   2.84 $\pm$ 0.04         \\
$\pi_{{\rm E},N}$              &  0.193  $\pm$ 0.008    &   0.210 $\pm$ 0.009       \\
$\pi_{{\rm E},E}$              &  0.211 $\pm$ 0.006     &   0.192 $\pm$ 0.008       \\
$ds/dt$ (yr$^{-1}$)            &  0.503 $\pm$ 0.049     &   0.467 $\pm$ 0.012       \\
$d\alpha/dt$ (rad yr$^{-1}$)   &  0.984 $\pm$ 0.068     &   -0.956 $\pm$ 0.090      \\
\hline                          
\end{tabular}
\tablecomments{${\rm HJD}^\prime={\rm HJD}-2450000$.\\
\smallskip}
\label{table:two}
\end{table}

\begin{figure}
\includegraphics[width=\columnwidth]{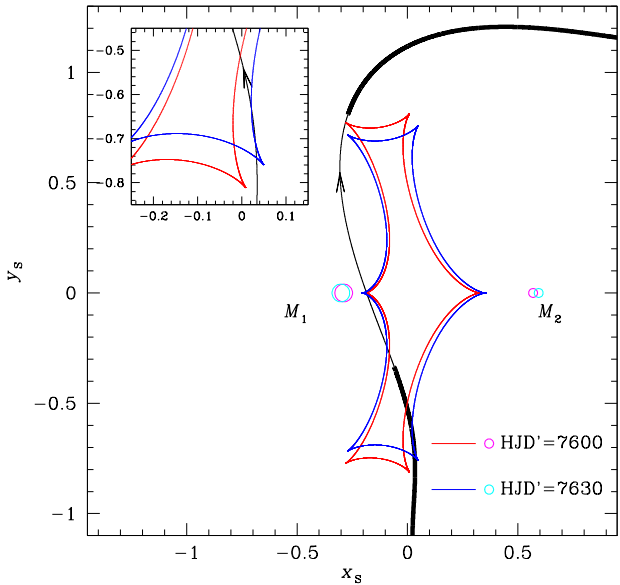}
\caption{
Configurations of the lens system. The curve with an arrow is the source trajectory and the 
cuspy closed figure represent the caustics. The small open circles marked by $M_1$ and $M_2$ 
represent the positions of the lens components.  Because the lens position and the caustic 
shape vary due to the orbital motion, we present lens positions and caustics at two epochs of 
${\rm HJD}^\prime=7600$ and $7630$. 
The part of the source trajectory marked in thin line represents the region
during which the bulge field was not observed as it passed behind the Sun.
The inset shows the zoom around the region of caustic entrance.
Coordinates are centered at the barycenter of the binary lens and lengths 
are scaled to the Einstein radius corresponding to the total mass of the lens. \\
\smallskip
}
\label{fig:five}
\end{figure}

In Figure~\ref{fig:five}, we present the lens system configuration in which the source 
trajectory (curve with an arrow) with respect to the caustic (closed curve with 6 cusps) 
and the individual lens components (open circles marked by $M_1$ and $M_2$) are shown.  The 
presented configuration is for the $u_0>0$ solutions. We note that the configuration of the 
$u_0<0$ solution is almost in mirror symmetry with respect to the $M_1$--$M_2$ axis compared 
to the $u_0>0$ solution. Because the lens positions and the resulting shape of the caustic 
vary in time due to the change of the binary separation caused by the lens-orbital effect, 
we present caustics at two epochs of ${\rm HJD}^\prime=7600$ (at the time of the bump) and 
${\rm HJD}^\prime=7630$ (at the time of the caustic entrance).  Due to the closeness of the 
binary separation to unity, the caustic forms a closed curve with 6 cusps and folds, 
i.e., a resonant caustic.  The configuration shows that the source approached close to the 
cusp located in the lower right part of the caustic producing the bump and passed over the 
adjacent fold of the caustic producing the caustic-crossing spike. The U-shape trough was 
produced during the passage of the source inside the caustic.  According to the best-fit model, 
the source exited the caustic by passing over the lower left fold at ${\rm HJD'}\sim 7695$ 
(2016-11-02).  Then, the source additionally passed the tip of the nearby cusp on 
${\rm HJD'}\sim 7715$ (2016-11-22) and approached the tip of the upper left cusp on 
${\rm HJD'}\sim 7778$ (2017-01-24), resulting in a multiple-peak light curve.  Unfortunately, 
these additional features were not be covered because the bulge could not be observed from 
Earth.

\begin{figure}
\includegraphics[width=\columnwidth]{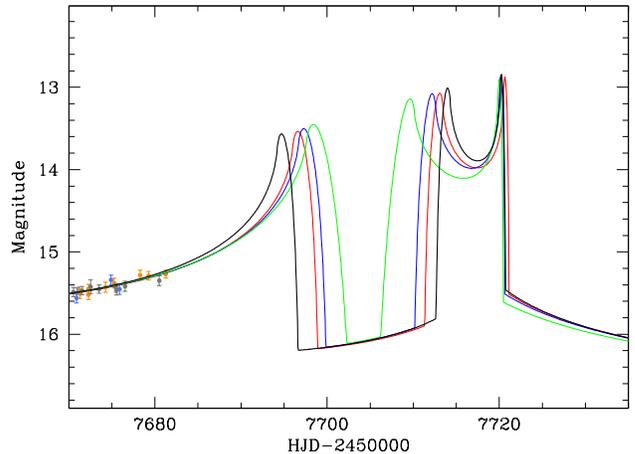}
\caption{
Variation of the unobserved part of the light curve. 
Presented are the model light curves for 4 different solutions within 
$3\sigma$ level from the best-fit solution (black curve). 
\smallskip
}
\label{fig:six}
\end{figure}

We note that the lensing parameters cloud have been better constrained if these additional 
caustic-related features could have been observed.  This is because these features are 
very sensitive to the small changes of the lensing parameters due to the special lens-system 
configuration in which the source approaches very close to the caustic.  The high sensitivity 
of the unseen caustic features to the lensing parameters is demonstrated in Figure~\ref{fig:six}, 
where we present model light curves for 4 different solutions within $3\sigma$ level from the 
best-fit solution (black curve).  It is found that the model light curve significantly varies 
even for slight differences in the lensing parameters.  It is known that multiple-peak features 
in lensing light curves help to better constrain lens systems \citep{An2001, Udalski2018}.  
If they had been observed, therefore, the lensing parameters, especially the higher-order 
parameters, could have been determined with improved precision and accuracy.

\begin{figure}
\includegraphics[width=\columnwidth]{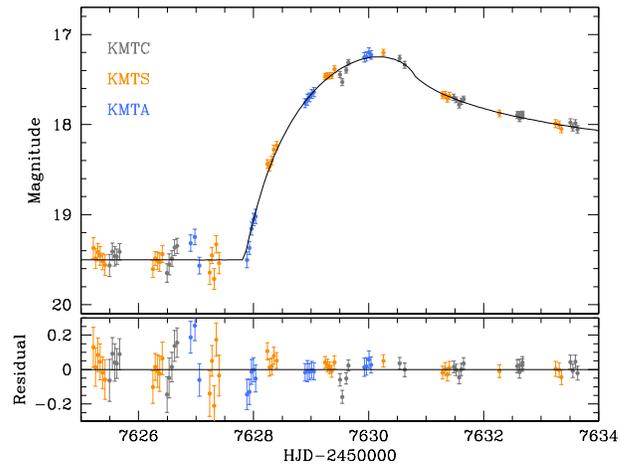}
\caption{
Zoom of the light curve around the time of the caustic crossing.
The curve plotted over data points is the model light curve.\\
\smallskip
}
\label{fig:seven}
\end{figure}

\begin{figure}
\includegraphics[width=\columnwidth]{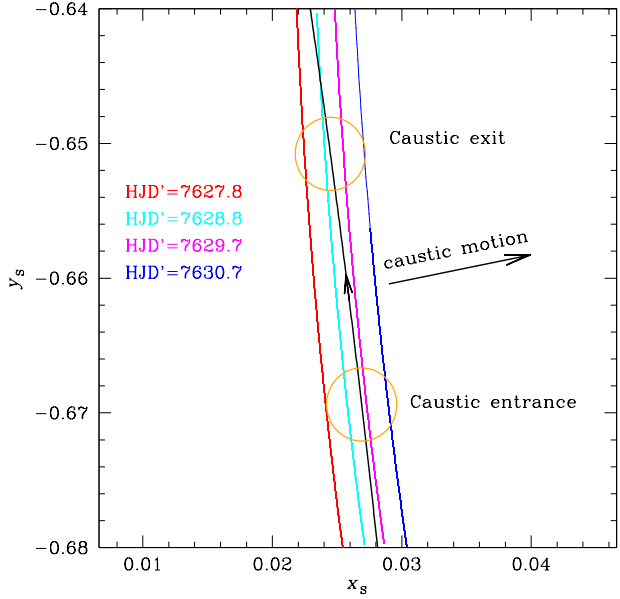}
\caption{
Zoom of the caustic configuration around the time of the caustic crossing.
The four curves in different colors represent fold caustics at different times marked in the legend
and the line with an arrow is the source trajectory.
The two orange circles represent the source at the beginning and end of the 
caustic crossing.\\
\smallskip
}
\label{fig:eight}
\end{figure}

Finite-source effects are clearly detected during the caustic crossing. In Figure~\ref{fig:seven}, 
we present the zoom of the light curve around the time of the caustic entrance. It shows that the 
crossing, which lasted for $\sim 3$ days, was densely and continuously covered from the combined 
observations using the globally distributed telescopes.  Analysis of this part of the light curve 
yields a normalized source radius of $\rho\sim 2.7\times 10^{-3}$ and a source self-crossing time 
scale of $t_*=\rho t_{\rm E}\sim 0.3$ days.  The duration between the time of the source star's 
touch to the fold of the caustic, at ${\rm HJD}_1^\prime=7627.8$, and the peak of the caustic 
crossing, at ${\rm HJD}_2^\prime=7630.2$, is $\Delta t\sim 2.4$ days.  For static caustic, this 
duration corresponds to $\Delta t=1.7(t_*/\sin\phi)$, where $\phi\sim 6^\circ$ is the angle 
between the source trajectory and the fold of the caustic \citep{Gould1999}.  Then, the apparent 
caustic-crossing time scale estimated from 
$\Delta t$ is $t_{*,{\rm app}}=(\sin \phi/1.7)\Delta t \sim 0.15$ days,  which is $\sim 2$ times 
shorter than the value estimated from $t_*=\rho t_{\rm E}\sim 0.3$ days. We find that the difference 
between $t_{*,{\rm app}}$ and $t_*$ is due to the movement of the caustic caused by the lens orbital 
motion.  This is shown  in Figure~\ref{fig:eight} where we present the zoom of the caustic 
configuration around the time of the caustic crossing.  It shows that the caustic moves rapidly 
toward right direction while the source moves slightly toward left direction.  This causes 
$t_{*,{\rm app}}$ shorter than $t_*$.

\section{Nature of the Lens}\label{section4}

\subsection{Angular Einstein Radius}\label{section4-1}

From the relations in Equations~(\ref{eq2}) and (\ref{eq3}), one needs two quantities of 
$\pie$ and $\thetae$ to uniquely determine the physical parameters of the lens mass and 
distance.  The microlens parallax is estimated from the measured microlens-parallax parameters 
by $\pie=(\pien^2+\piee^2)^{1/2}$. The angular Einstein radius is estimated from the measured 
normalized source radius by $\thetae=\theta_*/\rho$, where the $\theta_*$ is the angular source 
radius.  To determine the angular Einstein radius, then, it is required to estimate $\theta_*$.

\begin{figure}
\includegraphics[width=\columnwidth]{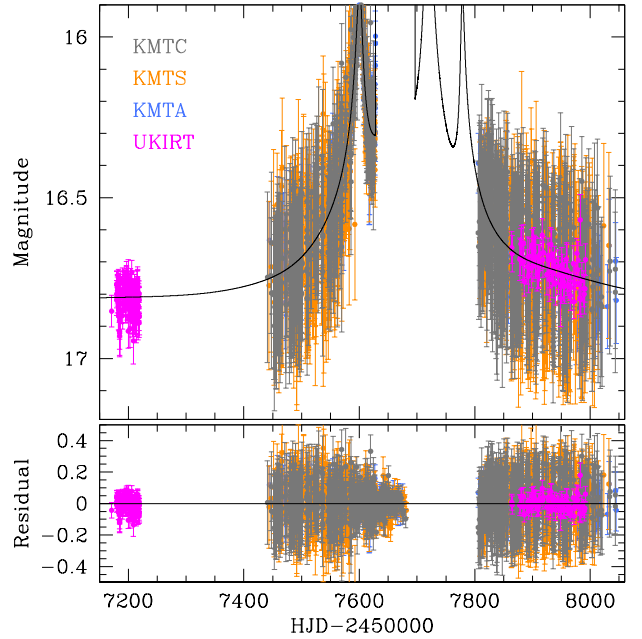}
\caption{
UKIRT data superposed by the model light curve.\\
\smallskip
}
\label{fig:nine}
\end{figure}

\begin{figure}
\includegraphics[width=\columnwidth]{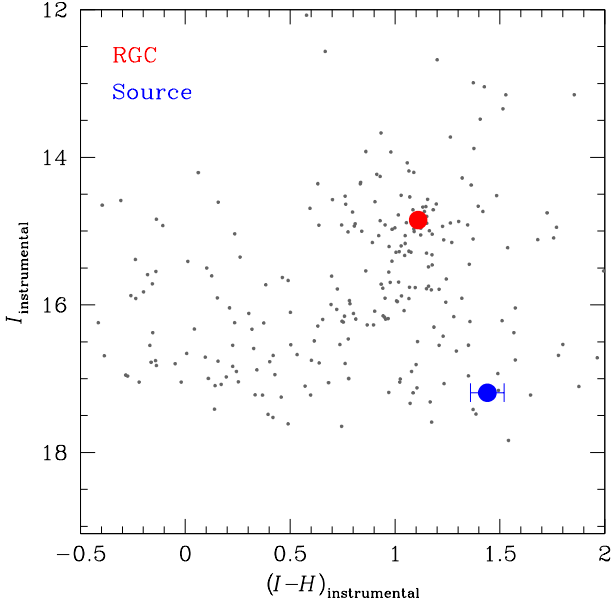}
\caption{
Source location with respect to the centroid of red giant clump (RGC) in the instrumental 
$(I-H,I)$ color-magnitude diagram. \\
\smallskip
}
\label{fig:ten}
\end{figure}

The angular source radius is estimated based on the de-reddened color and brightness.  To calibrate 
instrumental color and magnitude, we use the centroid of red giant clump (RGC), for which the color 
and brightness are known, as a reference \citep{Yoo2004}.  The measured instrumental $I$-band brightness 
of the source is $I=17.19\pm 0.02$, but the $V$-band brightness cannot be measured due to the poor 
photometry caused by severe extinction.  Instead of $V$-band photometry, we use $H$-band UKIRT data 
for the color measurement.  Figure~\ref{fig:nine} shows the UKIRT data superposed by the model curve.  
From model fitting, it is found that the $H$-band source brightness is $H=15.75\pm 0.08$ and thus 
$I-H=1.44\pm 0.08$.  To find the reference position of the RGC centroid, we construct an $(I-H,I)$ 
color-magnitude diagram by matching KMTC $I$-band and UKIRT $H$-band data.  Figure~\ref{fig:ten} 
shows the constructed $(I-H,I)$ color-magnitude diagram.  The position of the RGC centroid is 
$(I-H,I)_{\rm RGC}=(1.11,14.85)$.  From the known values of $(V-I, I)_{0,{\rm RGC}}=(1.06,14,43)$ 
\citep{Bensby2011, Nataf2013} and using the color-color relation \citep{Bessell1988}, the de-reddened 
$I-H$ color and $I$-band magnitude of the RGC centroid are  $(I-H, I)_{0,{\rm RGC}}=(1.29, 14.43)$.  
Combined with the measured offsets in color $\Delta(I-H)$ and magnitude $\Delta I$ of the source with 
respect to the RGC centroid, we find that the de-reddened color and brightness of the source are 
$(I-H,I)_0= (I-H, I)_{0,{\rm RGC}} + \Delta(I-H, I) = (1.62\pm 0.08,16.77\pm 0.02)$.  The measured 
$(I-H)_0$ color of the source corresponds to $(V-I)_0=1.51$.  This, combined with the brightness, 
indicates that the source is a K-type subgiant.  Once the de-reddened color of the source is determined, 
we then convert $V-I$ into $V-K$ using the color-color relation of \citet{Bessell1988} and then estimate 
$\theta_*$ using the relation between $V-K$ and the surface brightness \citep{Kervella2004}. From this 
procedure, we find that the angular source radius is $\theta_* = 3.25 \pm 0.34~\mu{\rm as}$.

\begin{table}
\centering
\caption{Einstein radius and relative proper motion.}
\begin{tabular}{lcc}
\hline\hline
\multicolumn{1}{c}{\ \ \ \ \ \ \ \ \ \ Parameter\ \ \ \ \ \ \ \ \ \ } &
\multicolumn{1}{c}{\ \ \ \ \ \ \ \ \ \ $u_0>0$\ \ \ \ \ \ \ \ \ \ }   &
\multicolumn{1}{c}{\ \ \ \ \ \ \ \ \ \ $u_0<0$\ \ \ \ \ \ \ \ \ \ }   \\
\hline
$\thetae$ (mas)                  &  1.20 $\pm$ 0.13  &  1.14 $\pm$ 0.12 \\
$\mu_{\rm geo}$ (mas yr$^{-1}$)  &  3.91 $\pm$ 0.42  &  3.71 $\pm$ 0.40 \\
$\mu_{\rm hel}$ (mas yr$^{-1}$)  &  2.82 $\pm$ 0.30  &  2.85 $\pm$ 0.31 \\
$\psi$                           &  $21^\circ$       &  $166^\circ$     \\ 
\hline
\end{tabular}
\label{table:three}
\end{table}

In Table~\ref{table:three}, we present the estimated angular Einstein radii for the $u_0>0$ 
and $u_0<0$ solutions. Also presented are the relative lens-source proper motions in the 
geocentric, $\mu_{\rm geo}$, and heliocentric frames, $\mu_{\rm hel}$. They are determined 
by \begin{equation}
\muvec_{\rm geo} = {\thetae \over t_{\rm E}}
{\pivec_{\rm E}\over \pie}
\end{equation}
\label{eq4}
and
\begin{equation}
\muvec_{\rm hel} = \muvec_{\rm geo} + {\bf v}_{\oplus,\perp} {\pi_{\rm rel}\over {\rm au}},
\label{eq5}
\end{equation}
respectively \citep{Gould2004, Dong2009}.  Here ${\bf v}_{\oplus,\perp}=(0.3,26.0)~{\rm km}~{\rm s}^{-1}$ 
represents the velocity of the Earth motion projected on the sky at $t_0$. The presented angle $\psi$ 
denotes the orientation angle of $\muvec_{\rm hel}$ as measured from the north.  We note that the 
measured value of the relative lens-source proper motion, $\mu_{\rm geo}\sim 3.9~{\rm mas}~{\rm yr}^{-1}$, 
is smaller than $\sim ~5~{\rm mas}~{\rm yr}^{-1}$ of typical lensing events.  Especially, the heliocentric 
proper motion, $\mu_{\rm hel}\sim 2.8~{\rm mas}~{\rm yr}^{-1}$, is nearly half of the typical value.  
We will discuss the cause of the slow relative lens-source motion in Section~\ref{section5}.

\begin{figure*}
\epsscale{0.85}
\plotone{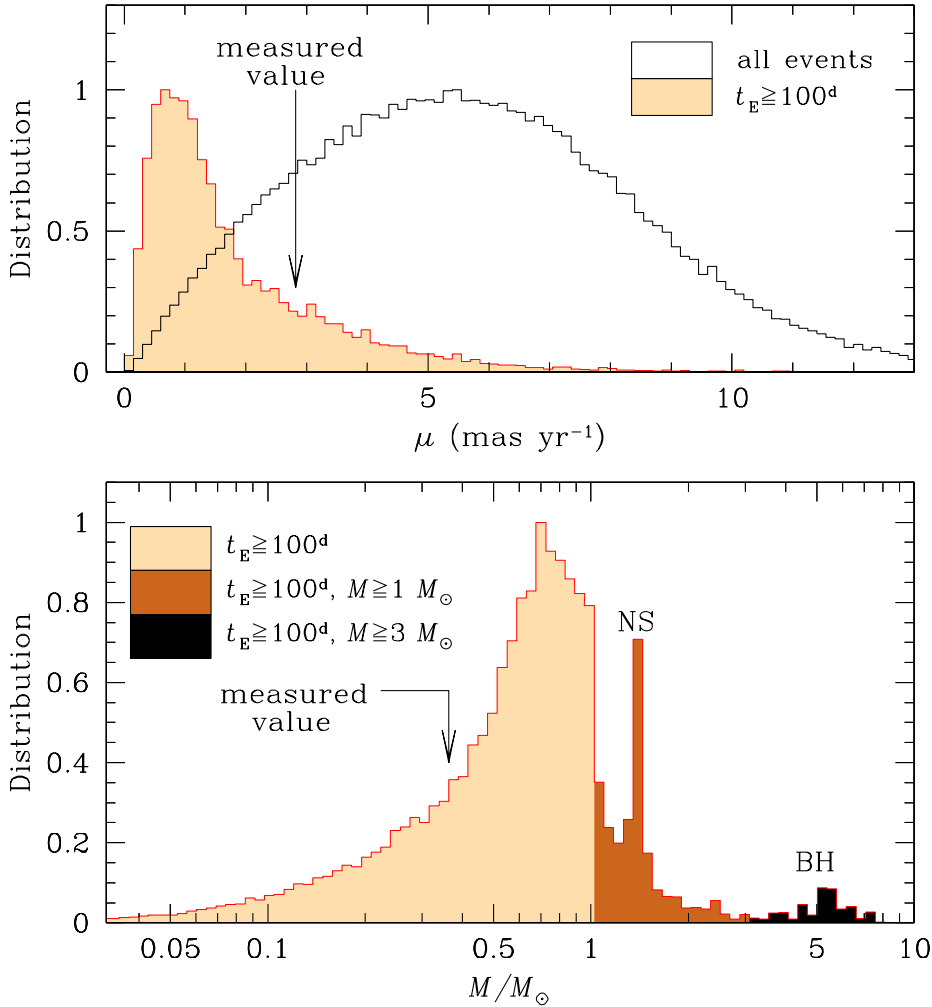}
\caption{
Distributions of the relative lens-source proper motion $\mu$ (upper panel) and lens mass 
$M$ (lower panel) for events with time scale $t_{\rm E}\geq 100$ days. For proper motions, 
we present two distributions where the shaded and unshaded distributions are for events 
$t_{\rm E}\geq 100$ days and for all events regardless of event time scales, respectively. 
For the lens mass distribution, the regions shaded by dark brown and black colors represent 
the distributions with lens masses $\geq 1.0~M_\odot$ and $\geq 3.0~M_\odot$, respectively. 
The values of $\mu$ and $M$ indicated by arrows represent the measured values of 
KMT-2016-BLG-2052L.
In the mass distribution, 
the peaks at $M\sim 1.35~M_\odot$ and $\sim 5~M_\odot$ are produced by 
neutron stars (NSs) and black holes (BHs), respectively.\\
\smallskip
}
\label{fig:eleven}
\end{figure*}

\subsection{Physical Parameters}\label{section4-2}

With the measured angular Einstein radius and the microlens parallax, we determine 
the mass and distance to the lens using the relations in Equations~(\ref{eq2}) and (\ref{eq3}). 
In Table~\ref{table:four}, we list the masses of the primary, $M_1$, and companion, $M_2$, 
of the lens, distance $D_{\rm L}$, and the projected separation between the lens components, 
$a_\perp=sD_{\rm L}\thetae$.  Also presented is the projected kinetic-to-potential energy 
ratio that is determined based on the total lens mass $M = M_1 +M_2$, the projected 
separation $a_\perp$, and the lens-orbital parameters $ds/dt$ and $d\alpha/dt$ by
\begin{equation}
\left( {{\rm KE}\over {\rm PE}} \right)_\perp =
{ (a_\perp/{\rm au})^3 \over 8\pi^2(M/M_\odot)}
\left[ \left( {1\over s}{ds/dt\over {\rm yr}^{-1}}\right)^2
+
\left( {d\alpha/dt\over {\rm yr}^{-1}} \right)^2\right].
\label{eq6}
\end{equation}
The lens system should meet the requirement $({\rm KE}/{\rm PE})_\perp\leq {\rm KE}/{\rm PE}\leq 1.0$ 
because otherwise the lens system would not be gravitationally bound. It is found that the ratios are 
$({\rm KE}/{\rm PE})_\perp\sim 0.45$ for both $u_0>0$ and $u_0<0$ solutions and the solutions meet the 
requirement.  This value is also in the expected range $0.2 \lesssim ({\rm KE}/{\rm PE})_\perp \lesssim 0.5$ 
for moderate eccentricity binaries that are not observed at unusual viewing angles.

\begin{table}
\centering
\caption{Physical lens parameters.}
\begin{tabular}{lcc}
\hline\hline
\multicolumn{1}{c}{\ \ \ \ \ \ \ Parameter\ \ \ \ \ \ \ } &
\multicolumn{1}{c}{\ \ \ \ \ \ \ $u_0>0$\ \ \ \ \ \ \ }   &
\multicolumn{1}{c}{\ \ \ \ \ \ \ $u_0<0$\ \ \ \ \ \ \ }   \\
\hline
$M_1$ ($M_\odot$)             &  0.34 $\pm$ 0.04    &  0.32 $\pm$ 0.04    \\
$M_2$ ($M_\odot$)             &  0.17 $\pm$ 0.02    &  0.17 $\pm$ 0.02    \\
$D_{\rm L}$ (kpc)             &  2.14 $\pm$ 0.20    &  2.23 $\pm$ 0.20    \\
$a_\perp$ (au)                &  2.45 $\pm$ 0.23    &  2.43 $\pm$ 0.22    \\
$({\rm KE}/{\rm PE})_\perp$   &  0.45               &  0.43               \\ 
\hline
\end{tabular}
\label{table:four}
\end{table}

The estimated masses of the lens components,
$M_1\sim 0.34~M_\odot$ for the primary and $M_2\sim 0.17~M_\odot$ for the companion, 
correspond to those of a mid and a late M-type main-sequence star, respectively.
Using the relation in Equation~(\ref{eq3}), the estimated distance to the lens is 
$D_{\rm L}\sim 2.1$~kpc. 
For the determination of $D_{\rm L}$, 
we use $\pi_{\rm S}={\rm au}/D_{\rm S}$ with 
the distance to the source estimated using the relation 
$D_{\rm S}=d_{\rm GC}/(\cos l + \sin l \cos\theta_{\rm bar}/\sin\theta_{\rm bar})\sim 8.06$ kpc,
where $d_{\rm GC}=8160$ pc is the Galactocentric distance, $\theta_{\rm bar}=40^\circ$ is the 
bulge bar orientation angle, and $l=0.58^\circ$ is the Galactic longitude of the source.
\citep{Nataf2013}.
The angular  Einstein radius is related to the distance to the lens by 
$\thetae\propto (D_{\rm L}^{-1}-D_{\rm S}^{-1})^{1/2}$ 
and thus the close distance to the lens results in the large angular Einstein radius. 
Because $t_{\rm E}=\thetae/\mu$, the long time scale of the event is 
caused by the combination of the slow relative lens-source motion and the 
large Einstein radius due to the close lens distance rather than the 
heavy mass of the lens.

\section{Discussion}\label{section5}

Because the event time scale is related to the lens mass and relative lens-source proper motion by
$t_{\rm E}\propto \sqrt{M}/\mu$, the long time scale of an event can be ascribed to either a large 
lens mass or a slow lens-source proper motion. For KMT-2016-BLG-2052, 
it turns out that the long time scale of the event is caused by 
the combination of the slow relative lens-source proper motion and the close 
distance to the lens rather than the heavy mass of the lens.
Then, a question is whether KMT-2016-BLG-2052 is an unusual case.  A related 
question is what the probability is for long time-scale events to be produced by very heavy lenses such 
as NSs and BHs.  In order answer these questions, we construct the probability distributions of 
relative lens-source proper motions and lens masses for long time-scale events by conducting Monte 
Carlo simulation of Galactic microlensing events.

The simulation is conducted based on the prior models of the matter density and dynamic distributions 
and the mass function of lens objects.  We adopt the \citet{Han2003} model for the matter density distribution.  
In this model, the disk and bulge follow a double-exponential distribution and a triaxial distribution, 
respectively.  The velocity distribution is based on  \citet{Han1995} model.  In the model, disk objects 
move following a gaussian distribution with a mean corresponding to the disk rotation speed and the 
motion of bulge objects follows a triaxial gaussian distribution with the velocity components along 
the axes determined based on the bulge shape using tensor virial theorem.  We use the initial mass 
function of \citet{Chabrier2003a} for the mass function of Galactic bulge objects and the present-day 
mass function of \citet{Chabrier2003b} for disk objects. We note that the adopted mass functions 
extend to substellar objects down to $0.01~M_\odot$. Because stellar remnants can cause long time-scale 
events, we include them in the mass function by assuming that stars with masses $1~M_\odot \leq 
M < 8~M_\odot$, $8~M_\odot \leq M < 40~M_\odot$, and $M \geq 40~M_\odot$ have evolved into white dwarfs 
(with a mean mass $\langle M\rangle\sim 0.6~M_\odot$), NSs (with $\langle M\rangle\sim 1.35~M_\odot$) 
and BHs (with $\langle M\rangle\sim 5~M_\odot$), respectively \citep{Gould2000a}.

In Figure~\ref{fig:eleven}, we present the distributions of the relative lens-source proper motions 
(upper panel) and lens masses (lower panel) for events with time scale $t_{\rm E}\geq 100$ days. To 
compare proper motions of long time-scale events with those of general events, we also present the 
proper-motion distribution for all events regardless of event time scales. From the comparison of 
the proper-motion distributions, one finds that long time-scale events tend to have substantially 
smaller proper motions, with a mode value of $\sim 0.8~{\rm mas}~{\rm yr}^{-1}$, than general events, 
with a mode $\sim 5~{\rm mas}~{\rm yr}^{-1}$.  The slow relative lens-source proper motion of long 
time-scale events is most likely caused by the chance alignment of the lens and source motion.  
Considering that the measured lens-source proper motion of KMT-2016-BLG-2052, 
$\mu_{\rm hel}\sim 2.8~{\rm mas}~{\rm yr}^{-1}$, 
is well within $2\sigma$ range of the distribution, the event is not an unusual case 
of long time-scale event.
From the distribution of lens masses, it is found 
that the probabilities that long time-scale events are produced by lenses with masses 
$\geq 1.0~M_\odot$ and $\geq 3.0~M_\odot$ are $\sim 19\%$ and 2.6\%, respectively.  This indicates 
that the majority of long time-scale events are produced by stellar lenses with masses 
$\lesssim 1.0~M_\odot$.  Considering that events produced by heavy lenses comprise a minor fraction 
of long time-scale events, it is essential to determine the lens mass by measuring both $\pie$ and 
$\thetae$ for the firm identification of stellar remnants such as NSs and BHs.

\section{Conclusion}

We analyzed the very long time-scale binary-lensing event KMT-2016-BLG-2052.  We revealed the 
nature of the lens by determining the lens mass from the simultaneous measurements of the microlens 
parallax and the angular Einstein radius.  The measured mass indicated that the lens was a binary 
composed of M dwarfs.  
We found that the long time scale of the event was caused by the combination of the slow relative 
lens-source motion and the large angular Einstein radius due to the close distance to the 
lens rather than the heavy mass of the lens.  
From the simulation of Galactic lensing events with very 
long time scales ($t_{\rm E}\gtrsim 100$ days), we found that long time-scale events tend to have 
substantially slow relative lens-source motions than general events.  We also found that the 
probabilities that long time-scale events were produced by lenses with masses $\geq 1.0~M_\odot$ 
and $\geq 3.0~M_\odot$ are $\sim 19\%$ and 2.6\%, respectively, indicating that events produced 
by heavy lenses comprise a minor fraction of long time-scale events.  The results indicate that 
it is essential to determine the lens masses by measuring both $\pie$ and $\thetae$ in order to 
firmly identify stellar remnants such as NSs and BHs.

\acknowledgements

Work by C.H.\ was supported by the grant (2017R1A4A1015178) of
National Research Foundation of Korea.
Work by A.G. was supported by JPL grant 1500811 and US NSF grant AST-1516842.
Work by J.C.Y.\ was performed under contract with
the California Institute of Technology (Caltech)/Jet Propulsion
Laboratory (JPL) funded by NASA through the Sagan
Fellowship Program executed by the NASA Exoplanet Science Institute.
This research has made use of the KMTNet system operated by the Korea
Astronomy and Space Science Institute (KASI) and the data were obtained at
three host sites of CTIO in Chile, SAAO in South Africa, and SSO in
Australia.


\begin{thebibliography}{}


\bibitem[Alard \& Lupton(1998)]{Alard1998} Alard, C., \& Lupton, R.~H.\ 1998, \apj, 503, 3
\bibitem[Albrow et al.(2009)]{Albrow2009} Albrow, M.~D., Horne, K., Bramich, D.~M., et al.\ 2009, \mnras, 397, 2099
\bibitem[Alcock et al.(1993)]{Alcock1993} Alcock, C., Akerlof, C.~W., Allsman, R.~A., et al. 1993, Nature, 365, 621
\bibitem[An \& Gould(2001)]{An2001} An, J.~H., \& Gould, A.\ 2001, \apj, 563, L111
\bibitem[Aubourg et al.(1993)]{Aubourg1993} Aubourg, E., Bareyre, P., Br\'ehin, S., et al.\ 1993, Nature, 365, 623
\bibitem[Batista et al.(2011)]{Batista2011} Batista, V., Gould, A., Dieters, S., et al.\ 2011, \aap, 529, 102
\bibitem[Bensby et al.(2011)]{Bensby2011} Bensby, T., Ad\'en, D., Mel\'endez, J., et al.\ 2011, \pasp, 533, 13
\bibitem[Bessell \& Brett(1988)]{Bessell1988} Bessell, M.~S., \& Brett, J.~M.\ 1988, \pasp, 100, 1134
\bibitem[Bond et al.(2001)]{Bond2001} Bond, I.~A., Abe, F., Dodd, R.~J., et al.\ 2001, \mnras, 327, 868
\bibitem[Chabrier(2003a)]{Chabrier2003a} Chabrier, G.\ 2003a, \pasp, 115, 763
\bibitem[Chabrier(2003b)]{Chabrier2003b} Chabrier, G.\ 2003b, \apjl, 586, L13
\bibitem[Dong et al.(2009)]{Dong2009} Dong, S., Gould, A., Udalski, A., et al.\ 2009, \apj, 695, 970
\bibitem[Gould(1992)]{Gould1992a} Gould, A.\ 1992, \apj, 392, 442
\bibitem[Gould(2000a)]{Gould2000a} Gould, A.\ 2000a, \apj, 535, 928
\bibitem[Gould(2000b)]{Gould2000b} Gould, A.\ 2000b, \apj, 542, 785
\bibitem[Gould(2004)]{Gould2004} Gould, A.\ 2004, \apj, 606, 31
\bibitem[Gould \& Loeb(1992)]{Gould1992b} Gould, A., \& Loeb, A.\ 1992, \apj, 396, 104
\bibitem[Gould \& Andronov(1999)]{Gould1999} Gould, A., \& Andronov, N.\ 1999, \apj, 516, 236
\bibitem[Han \& Gould(1995)]{Han1995} Han, C., \& Gould, A.\ 1995, \apj, 447, 53
\bibitem[Han \& Gould(2003)]{Han2003} Han, C., \& Gould, A.\ 2003, \apj, 592, 172
\bibitem[Han et al.(2016)]{Han2016} Han, C., Udalski, A., Lee, C.-U., et al.\ 2016, \apj, 827, 11
\bibitem[Kervella et al.(2004)]{Kervella2004} Kervella, P., Th\'evenin, F., Di Folco, E., \& S\'egransan, D. 2004, \aap, 426, 29
\bibitem[Kim et al.(2018a)]{Kim2018a} Kim, D.-J., Kim, H.-W., Hwang, K.-H., et al.\ 2018a, \aj, 155, 76
\bibitem[Kim et al.(2018b)]{Kim2018b} Kim, H.-W., Hwang, K.-H., Kim, D.-J., et al.\ 2018b, \aj, 155, 186
\bibitem[Kim et al.(2016)]{Kim2016} Kim, S.-L., Lee, C.-U., Park, B.-G., et al.\ 2016, JKAS, 49, 37
\bibitem[Mao \& Paczynski(1991)]{Mao1991} Mao, S., \& Paczynski, B.\ 1991, \apj, 374, L37
\bibitem[Nataf et al.(2013)]{Nataf2013} Nataf, D.~M., Gould, A., Fouqu\'e, P., et al.\ 2013, \apj, 769, 88
\bibitem[Paczy\'nski(1986)]{Paczynski1986} Paczy\'nski, B.\ 1986, \apj, 304, 1
\bibitem[Schechter et al.(1993)]{Schechter1993} Schechter, P.~L., Mateo, M., \& Saha, A.\ 1993, \pasp , 105, 1342
\bibitem[Shvartzvald et al.(2015)]{Shvartzvald2015} Shvartzvald, Y., Udalski, A., Gould, A., et al.\ 2015\ apj, 814, 111
\bibitem[Shvartzvald et al.(2017)]{Shvartzvald2017} Shvartzvald, Y., Bryden, G., Gould, A., et al.\ 2017, \aj, 153, 61 
\bibitem[Skowron et al.(2011)]{Skowron2011} Skowron, J., Udalski, A., Gould, A., et al. 2011, \apj, 738, 87
\bibitem[Smith et al.(2003)]{Smith2003} Smith, M. C., Mao, S., \& Paczy \'nski, B.\ 2003, \mnras, 339, 9
\bibitem[Udalski(1993)]{Udalski1993} Udalski, A., Szyma\'nski, M., Ka{\l}u\.zny, J., Kubiak, M., Krzemi\'nski, W., 
    Mateo, M., Preston, G.~W., \& Paczy\'nski, B.\ 1993, Acta Astron., 43, 289
\bibitem[Udalski et al.(2018)]{Udalski2018} Udalski, A., Han, C., Bozza, V.\ 2018, \apj, 853, 70
\bibitem[Udalski et al.(2015)]{Udalski2015} Udalski, A., Szyma\'nski, M. K., \& Szyma\'nski, G.\ 2015, AcA, 65, 1
\bibitem[Wo\'zniak(2000)]{Wozniak2000} Wo\'zniak, P.~R.\ 2000, AcA, 50, 42
\bibitem[Wyrzykowski et al.(2016)]{Wyrzykowski2016} Wyrzykowski, {\L}., Kostrzewa-Rutkowska, Z., Skowron, J., et al.\ 2016, \mnras, 458, 3012
\bibitem[Yoo et al.(2004)]{Yoo2004} Yoo, J., DePoy, D.~L., Gal-Yam, A., et al.\ 2004, \apj, 603, 139


\end{thebibliography}
\end{document}